\documentclass[amsmath, amssymb,showpacs, aps,prb,reprint,superscriptaddress]{revtex4-1}
\usepackage{graphicx} 
\usepackage{dcolumn}
\usepackage{bm}
\usepackage[applemac]{inputenc}

   \usepackage{hyperref}

\usepackage{enumerate}
\begin{document}


\title{Correlation Between Local Elastic Heterogeneities and Overall Elastic Properties in Metallic Glasses }


\author{B. A. Sun}
\email{ifwsun@gmail.com}
\affiliation{Centre for Advanced Structural Materials, Department of Mechanical and \\Biomedical Engineering, City University of Hong Kong, Kowloon, Hong Kong}

\author{Y. C. Hu}
\affiliation{Centre for Advanced Structural Materials, Department of Mechanical and \\Biomedical Engineering, City University of Hong Kong, Kowloon, Hong Kong}
\affiliation{Institute of Physics, Chinese Academy of Science, 100190, Beijing, China}

\author{D. P. Wang}
\affiliation{Centre for Advanced Structural Materials, Department of Mechanical and \\Biomedical Engineering, City University of Hong Kong, Kowloon, Hong Kong}
\author{P. Wen}
\affiliation{Institute of Physics, Chinese Academy of Science, 100190, Beijing, China}

\author{W. H. Wang}
\affiliation{Institute of Physics, Chinese Academy of Science, 100190, Beijing, China}
\author{C. T. Liu}
\affiliation{Centre for Advanced Structural Materials, Department of Mechanical and \\Biomedical Engineering, City University of Hong Kong, Kowloon, Hong Kong}

\author{Y. Yang}
\email{yonyang@cityu.edu.hk}
\affiliation{Centre for Advanced Structural Materials, Department of Mechanical and \\Biomedical Engineering, City University of Hong Kong, Kowloon, Hong Kong}


\date{\today}

\begin{abstract}
The common notion suggests that metallic glasses (MGs) are a homogeneous solid at the macroscopic scale; however, recent experiments and simulations indicate that MGs contain nano-scale elastic heterogeneities. Despite the fundamental importance of these findings, a quantitative understanding is still lacking for the local elastic heterogeneities intrinsic to MGs. On the basis of Eshelby's theory, here we develop a micromechanical model that correlates the properties of the local elastic heterogeneities, being very difficult to measure experimentally, to the measurable overall elastic properties of MGs, such as shear/bulk modulus and Poisson's ratio. Our theoretical modeling is verified by the experimental data obtained from various MGs annealed to different degrees. Particularly, we revealed that the decrease of Poisson's ratio upon annealing of MGs is associated with a much large shear softening over hydrostatic-pressure softening, and $vice$ $versa$ in local elastic inhomogeneities. The relative extent of the bulk versus shear modulus softens is extracted for different MGs, and is found to closely depend on the specific composition and their ductility. The implication of our results on the Poisson's ratio criterion on the ductility as well as the aging dynamics in MGs is discussed.
\end{abstract}

\pacs{61.82.Bg,62.20.de, 62.20.dj, 62.20.fk}

\maketitle
\section{\label{I}Introduction}
Despite the fundamental and technological importance, the structure-property relationship in metallic glasses (MGs) has long been obscured by their long-range disordered structure\cite{ChenmwAnnuRew,Ma:2015aa,Schroer}. A MG often looks microstructurally featureless or ''amorphous'' under a conventional electron microscope, which is in sharp contrast to crystalline alloys in which grain boundaries, dislocations and many other microstructural features can be readily identified\cite{Ma:2015aa}. However, recent atomistic simulations \cite{Patrick-Royall:2008aa,Dingpnas,Cheng2011379}and experiments\cite{Egami2010,LiuYHPRL2011,Wagner:2011aa,Ye:2010aa} indicate that the amorphous structure of glasses is indeed heterogeneous, containing nano-scale elastic\cite{Egami2010,Wagner:2011aa} or viscoelastic\cite{LiuYHPRL2011,Ye:2010aa} imhomogeneities on top of their intrinsic density fluctuation. Unlike the conventional microstructural defects in crystalline alloys, the structural heterogeneities in MGs are dynamic in nature, which are invisible under static electron microscopes but could be theoretically associated with low-frequency vibration modes\cite{Dingpnas,MazzaEPL1996,Schoberprb} or liquid-like regions with very low local viscosities or relaxation times\cite{Ye:2010aa,Jiaow2013}. In principle, these regions of dynamic heterogeneity are susceptible to irreversible local atomic rearrangements under external perturbations (stress or heat), thus acting as ''flow units" to initiate macroscopic plastic flows\cite{Ke2014560,Krisponeit:2014aa}or as "liquid-like sites" to initiate secondary relaxations in MGs\cite{Wang:2014ab,Wang:2015aa}. Despite the recent efforts confirming the existence of the dynamic heterogeneities in MGs, however, it still remains challenging to derive the mechanical/physical properties of MGs, in a quantitative manner, from the perspective of local structural heterogeneities.      
    On the other hand, through the extensive studies over the past decades\cite{Wang2012487,Sun2015211}, various correlations were established between the overall elastic constants of MGs and their many other physical properties. Today, it is commonly thought that the elastic properties of MGs may hold the key to the understanding of many fundamental issues in MGs, such as glass transition\cite{DyreRevModPhys,Novikov:2004aa,Egami2006882}, relaxation\cite{EvensonPRB,Huo20134329}and deformation\cite{Lewandowski2005,NgaiJCP}. For example, it was shown that the elastic modulus of a MG is determined by both the atomic bonding strength and the atomic configuration\cite{ChengPhysRevB,Huo20134329}, which varies with the possible structural change in a MG, as witnessed in a typical structural relaxation\cite{Lewandowski2005,AroraJAP,KURSUMOVIC19801303} or rejuvenation process\cite{Ketov:2015aa,Tong2015240}. Moreover, extensive studies also indicate that the Poisson's ratio of a MG correlates with the structural state of the corresponding supercooled liquid, as characterized by the liquid fragility\cite{Novikov:2004aa}, which in turn affects the ductility of the glass. As motivated by the fact that the amorphous structure of MGs is overall heterogeneous, we intend to derive relations in this work that connects the elastic constants of MGs to their local structural heterogeneities. Once these relations become available, one may develop important structure-property relations, with the use of the elastic constants as intermediate variables, which ultimately bridge the structural heterogeneity in MGs and their physical and mechanical properties or other attributes. 

\section{\label{II}Theoretical modelling}
\begin{figure}
 \centering
\includegraphics{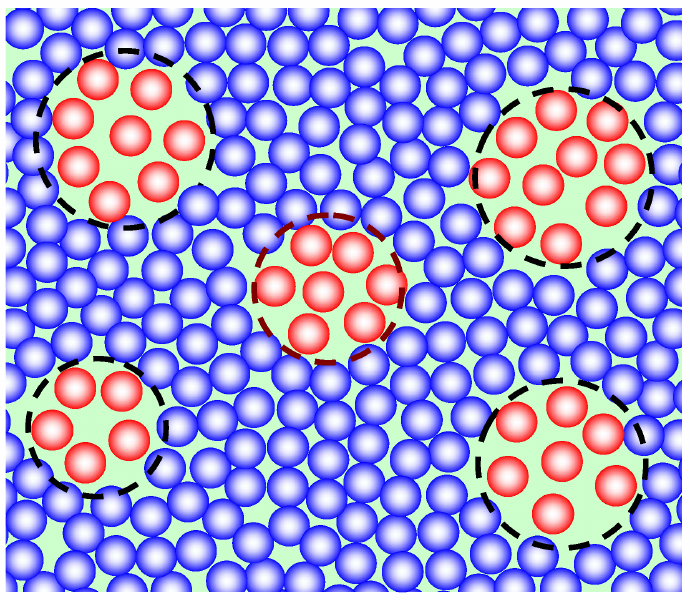}
\caption{\label{fig1} The schematic for the atomic-scale structural heterogeneities in MGs. The loosely packed region (in red atoms) are embedded in the matrix of densely-packed region (in blue atoms).}
\end{figure}

In principle, the whole structure of an MG can be envisaged as an atomic-scale composite\cite{Ye:2010aa,Liu201376}, consisting the solid-like regions (SLRs) as a "matrix" and isolated soft regions as "inclusions", as schematically shown in Figure \ref{fig1}. The LLRs are loosely packed atom regions, and should have the lower elastic moduli (bulk or shear modulus) than those of rigid SLRs. From the point view of mechanics, LLRs can be viewed as elastic imhomogeneities embedded in an isotropic matrix. Thus, the effective elastic moduli of the whole structure can be readily obtained from the Eshelby's theory\cite{Eshelby} on inhomogeneities. The Eshelby's theory have been extensively used in analyzing the elastic stress field associated with the shear transformation of elementary deformation units\cite{LangerPhysRevE.64.011504,MaloneyPhysRevLett.93.016001,ProcacciaPRL}, which are possibly initiated from LLRs. Here, we consider the static effective elastic modulus of MGs from the Eshelby's approach on the elastic inhomogeneities. Suppose that the composite structure is subject to a uniform applied elastic strain field of $\epsilon_{ij}^{A}$($i$,$j$ over 1,2,3) by the surface traction, the elastic mismatch between inhomogeneities and the matrix will cause additional stress/strain field, thus leading to an interaction elastic energy. The effective elastic modulus ($C_{ijkl}$) is defined from the total elastic energy of the specimen:
                           \begin{equation}\label{eq1}
 E_{total}=E_{0}+\sum{E_{int}}=\frac{1}{2}C_{ijkl}\epsilon_{ij}^{A}\epsilon_{kl}^{A}
                            \end {equation}
In Eq.\ref{eq1}, $E_{0}$ is the elastic energy of the specimen under a uniform $\epsilon_{ij}^{A}$  when it is free of inhomogeneities. $E_{int}$ is the interaction energy between an inhomogeneity and the external field. The sum is over all inhomogneities contained in the matrix. Both $E_{0}$ and $E_{int}$ are the quadratic functions of $\epsilon_{ij}^{A}$. Assume the bulk (or shear) modulus of SLRs and LLRs are $K_{0}$ (or $G_{0}$) and $K_{1}$ (or $G_{1}$), respectively, the effective bulk ($K$) or shear ($G$) modulus for the SSRs-LLRs assembly is derived as (See the theoretical analysis in Supplementary Text I):
             \begin{equation}\label{eq2}
 K=\frac{K_{0}}{1+AV_{f}}, G=\frac{G_{0}}{1+BV_{f}}
                             \end {equation}
 where $V_{f}$ is the volume fraction of LLRs, $K_{0}$ and $G_{0}$ should have values close to that of ideal glasses. $A$ and $B$ are two coefficients associated with the discrepancy of elastic moduli between SLRs and LLRs. It should be noted here that the elastic moduli of LLRs in real structure of MGs are not uniform and might have a distribution spectrum, however, the current theoretical approach is still valid from a mean-field sense if $K_{1}$ or $G_{1}$ is viewed as the mean value of bulk modulus and shear modulus of LLRs distributed in MGs, respectively.  For spherical LLRs, $A $ and $B$ read\cite{Eshelby}:
               \begin{equation}\label{eq3}
A=\frac{K_{1}-K_{0}}{(K_{0}-K_{1})\alpha-K_{0}}, B=\frac{G_{1}-G_{0}}{G_{0}-G_{1})\beta-G_{0}}
                             \end {equation}
where $\alpha$ and $\beta$ are two parameters depending on the Poisson's ratio of the matrix, $\nu_{0}$: $\alpha=(1/3)(1+\nu_{0})/(1-\nu_{0})$, $\alpha=(2/15)(4-5\nu_{0})/(1-\nu_{0})$. $A$ and $B$ usually have positive values due to elastic softening of LLRs compared with SLRs ($K_{1}<K_{0}$, $G_{1}<G_{0}$). Finally, the effective Poisson's ratio of MGs can be derived from Eq.\ref{eq2} based on relations of elastic constants in an isotropic material:
                          \begin{equation}\label{eq4}
\nu=\frac{(1+\nu_{0})(1+BV_{f})-(1-2\nu_{0})(1+AV_{f})}{2(1+\nu_{0})(1+BV_{f})+(1-2\nu_{0})(1+AV_{f})}
                          \end {equation}    
where $\nu_{0}=(3K_{0}-2G_{0})/(6K_{0}+2G_{0})$, is the Poisson's ratio of SLR matrix. 

\begin{figure}
 \centering
\includegraphics[width=8cm]{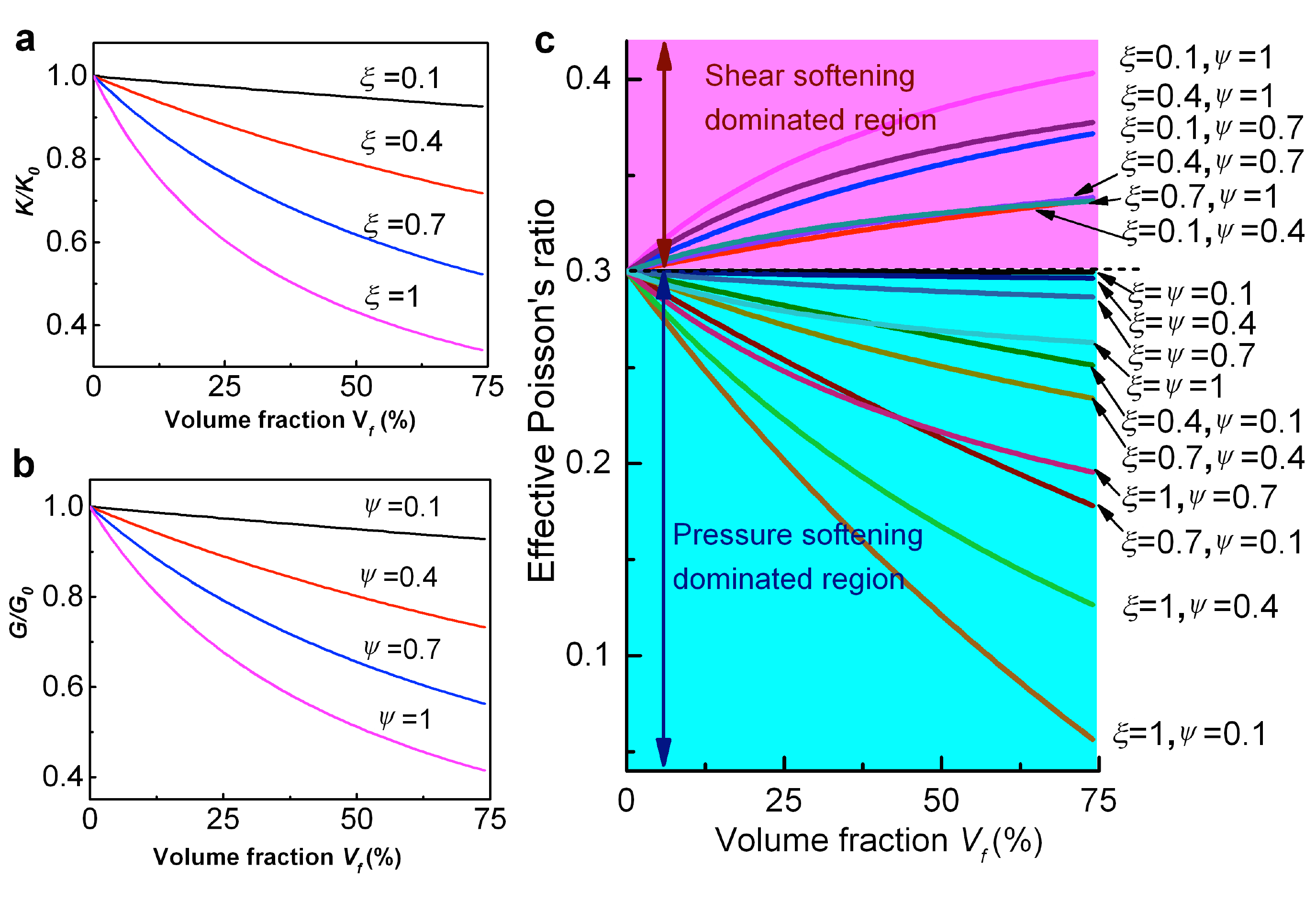}
\caption{\label{fig2} The calculated variation of effective elastic constants with the volume fraction of LLRs for different values of $\xi$ and $\psi$: (a) the effective bulk modulus $K$; (b) the effective shear modulus $G$; (c) the effective Poisson's ratio $\nu$. }
\end{figure}

 Eqs. \ref{eq2}-\ref{eq4} in fact establish general relations between the macroscopic elastic properties and the characteristics of microscopic structural heterogeneities in MGs. From these relations, one can generally found that the overall elastic moduli of MGs do not follows the conventional "rule of mixtures", i.e. the weighted average of elastic moduli of SLRs and LLRs. This is unexpected since the rule is often commonly used in previous studies when estimating the elastic properties of MGs or composite materials\cite{Wang2012487}. To facilitate discussion, we define $\xi=(K_{0}-K_{1})/K_{0}$, $\psi=(G_{0}-G_{1})/G_{0}$, which reflect the elastic softening degree of LLR compared with SLRs. In general, $\xi$ and $\psi$, range from 0 to 1. In limiting cases, for $\xi=1$, 
 $\psi=1$, LLRs become cavitation, while for $\xi<1$, $\psi=1$, LLRs become normal liquids with a Poisson's ratio approaching 0.5\cite{Greaves:2011aa}. Thus, Eq.\ref{eq4} in general formulate the overall Poisson's ratio when a solid matrix containing spherical inhomogeneities in various matter states from cavitation, liquids to glasses. Figure \ref{fig2} shows variations of overall elastic moduli with the volume fraction of SLRs at different values of $\xi$ and $\psi$. As can be seen, both $K$ and $G$ decrease monotonically with $V_{f}$. The extent that $K$ or $G$ decrease at a certain $V_{f}$ depends on the specific value of $\xi$ or $\psi$: The larger the $\xi$ or $\psi$ is, the more steeply the $K$ or $G$ drops. While the Poisson's ratio could either decrease or increase with $V_{f}$, depending on specific values of $\xi$ and $\psi$. Based on this, one can define a shear-softening dominated region and a pressure-softening dominated region, as shown in Figure \ref{fig2}c. In the shear-softening dominated region, $\xi < \psi$, $\nu$ increase monotonically with $V_{f}$. This agree with the common notion that introducing more LLRs regions into the glassy structure will lead to a higher Poisson's ratio, thus promoting the plasticity in MGs\cite{Li:2015aa,Wang:2014ab}. In contrast, in the pressure-softening region, $\xi\ge\psi$ , $\nu$ decrease monotonically with $V_{f}$. This case has not been reported in previous studies. However, we will show experimental evidences for the case from structural relaxation of a Ce-based MG below. Depending on the value of $\nu_{0}$, $\nu$ could even reach a negative value at large $V_{f}$ and $\xi-\psi$, resulting in materials with negative Poisson's ratio\cite{Greaves:2011aa}.

\section{\label{III}Experimental Methods and Results}

\subsection{\label{BI}Experimental Methods}

MG alloy ingots with the nominal compositions Zr$_{52.5}$Cu$_{17.9}$Al$_{10}$Ni$_{14.6}$Ti$_{5}$(Vit105), Zr$_{41}$Ti$_{14}$Cu$_{12.5}$Ni$_{10}$Be$_{22.5}$(Vit1), Zr$_{46.75}$Ti$_{8.25}$Cu$_{7.5}$Ni$_{10}$Be$_{27.5}$(Vit4) and Ce$_{68}$Al$_{10}$Cu$_{20}$Fe$_{2}$ were produced by arc melting a mixture of pure metals (purity $\ge$99.5\% in mass weight) in a Ti-gettered argon atmosphere. Glassy alloy rods with diameters of 2-10 mm and a length of at least 30 mm were obtained by suction casting into a water-cooling copper mould. The amorphous nature of specimens both at as-cast state and after thermal annealing were confirmed by the x-ray diffraction (XRD) method using a MAC Mo3 XHF diffractometer with Cu K$\alpha$ radiation and the differential scanning calorimetry (DSC, Perkin Elmer DSC7). Thermal annealing experiments were performed in a Muffle furnace. MG samples were first encapsulated into a quartz tube with a vacuum of $10^{-4}$ Pa and then annealed at their sub-$T_{g}$ temperatures for different times. The annealing temperature for Vit105, Vit1, Vit4 and the Ce-based MGs is 600 K, 623 K, 523 K and 333 K, respectively. After a certain annealing time, the glassy sample was removed from the furnace and cooled to room temperature for elastic modulus measurement. Then the same sample is encapsulated again and put back to the furnace for further annealing. This process was repeated until all elastic measurements are completed at different annealing times. 

The elastic modulus were measured at room temperature by using a pulse echo overlap method \cite{WenPhysRevB.73.224203,Wang2012487}with a MATEC 6600 ultrasonic system. The frequency of the ultrasonic wave was 10 MHz. The acoustic longitudinal velocity, $v_{l}$, and shear velocity, $v_{s}$, of MG samples were measured at room temperature  The density $\rho$ was measured by Archimedes' principle in distilled water in an accuracy of 0.5\%. The elastic constants, e.g. bulk modulus $K$, shear modulus $G$, Young's modulus $E$ and Poisson's ratio of MGs are derived from acoustic velocities\cite{Wang2012487}. The uniaxial compression tests were performed on an Instron 5869 electromechanical test system under a constant strain rate $5\times10^{-4}$ s$^{-1}$. The load, displacement and the time are recorded at a frequency of 50 Hz. The strain was measured by a laser extensometer(Fiedler) attached to the testing machine.  

MD simulations for five MGs, i.e. Cu$_{50}$Zr$_{50}$, Cu$_{46}$Zr$_{46}$Al$_{8}$, Mg$_{65}$Cu$_{25}$Y$_{10}$, Ni$_{33}$Zr$_{67}$, Pd$_{82}$Si$_{18}$, were performed using the code LAMMPS. The embedded-atom method (EAM) potentials were employed to describe the interatomic interactions\cite{Cheng2011379}. For each model, with periodic boundary conditions applied in three dimensions, the initial configuration containing 16,000 atoms was first equilibrated at 2000 K for 1.5 ns, followed by rapid quenching (1012 K/s) to 300 K in NPT (constant number, constant pressure and constant temperature) ensemble.  To extract the bulk moduli of metallic glasses at various temperatures, an equal deformation strain   is applied along the three axes of the simulation box, resulting in a volumetric change. The corresponding pressure p is measured. The bulk modulus is determined from the slope of  pressure-dilation strain plots. In all of the simulations, the time step used to integrate the equations of motion was set to be 0.001 ps.

\subsection{\label{BII}The variation of elastic moduli with time during structural relaxation}
\begin{figure}
 \centering
\includegraphics[width=8cm]{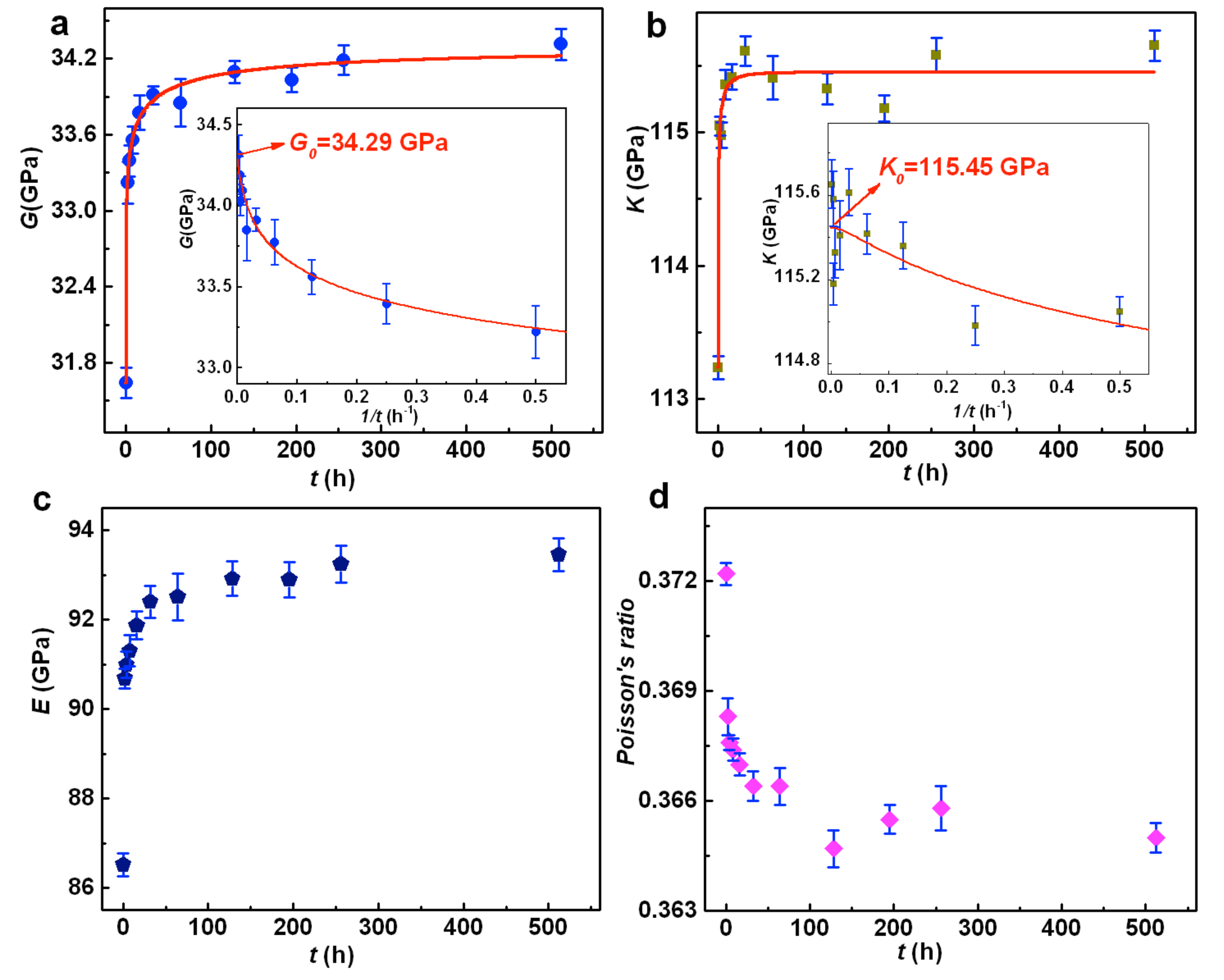}
\caption{\label{fig3} The change of measured elastic constants with the annealing time $t$, respectively for Vit105 MG: (a) bulk modulus $K$; (b) shear modulus $G$; (c) Young's modulus $E$; (d) Poisson's ratio $\nu$. The inset shows variations of $K$ and $G$ with the inverse time, so $K_{0}$ and $G_{0}$ can be easily identified. }
\end{figure}

We annealed MGs of different compositions (listed in Table \ref{tableI}) under sub-$T_{g}$  temperatures for various times and then measured the variation of their elastic properties by a pulse echo overlap ultrasonic method. After long-time annealing, no crystallization occurs and the sample still remains a fully amorphous structure (see Supplementary Figure 1). During annealing, one expects that the volume fraction of LLRs in the glassy structure will decrease generally with time due to the structural-relaxation induced densitified process, yet the chemical composition of MGs remain unchanged. Thus, the correlation between elastic properties and microscopic structural heterogeneities in MGs can be tracked. Figure \ref{fig3} shows the change of measured elastic constants ($K$, $G$, $E$, $\nu$) with the annealing time $t$, respectively for a typical Vit105 MG. In accordance with previous studies, one can see that $K$, $G$ and $E$ increase sharply at the initial stage and then gradually approach a saturation plateau at long time, while v continuously decrease with the time. The $K$ or $G$ can be well fitted by:
                             \begin{equation}\label{eq5}
P=\frac{P_{\infty}}{1+c_{0}exp(-kt^{n})}
                          \end {equation}   
 where $P$ can be either $K$ or $G$. $P_{\infty}$, $c_{0}$, $n$ is the constant parameters. $E$ and $\nu$ are not fitted here since they depend on $K$ and $G$. The variation of $K$ or $G$ with time for other MGs displayed similar trends with Vit105, and can be fitted by Eq.\ref{eq5}. The fitting values of parameters are listed in Table \ref{tableI}. One can see that the exponent $n$, and ranges from 0 to 1, depending on the specific MG composition. The $K_{\infty}$ and $G_{\infty}$ are the modulus upon infinite-time annealing, and can be regarded to equal those of SLRs, $K_{0}$ and $G_{0}$, respectively. When the term $c_{0}exp(-kt^{n})$ is much smaller than 1, for the first-order approximation, Eq.\ref{eq5} is reduced to $P_{\infty}[1-exp(-kt^{n})]$, in good agreement with the formula used to describe the kinetics of changes in elastic properties on annealing different kinds of glasses in previous studies.\cite{NemilovPM,DaviesAdvPhys} 

 \begin{table*}
 \caption{\label{tableI}The values of various parameters obtained by fitting the elastic modulus data during structural relaxation (bulk modulus $K_{0}$, shear modulus $G_{0}$ and Poisson's ratio $\nu_{0}$ of elastic matrix, the exponent $n$, the ratio $A/B$, the relative softening of $\psi/\xi$) and the plastic strain $\epsilon_{p}$ for various MGs.}
\begin{ruledtabular}
\begin{tabular}{lcccccccc}
 MGs &$K_{0}$(GPa)&$G_{0}$(GPa)&$\nu_{0}$&$n$
 &$A/B$ &$\psi/\xi$ &$\psi$(\%)&$\epsilon_{p}$\\
\hline
Vit105& 115.48 & 34.29 & 0.365 & 0.24
& 0.193 & 4.5 & 45 & 5.5 \\
Vit1& 115.42 & 43.56 & 0.332 &0.72
& 0.146 & 5.4 & 54 & 3.3 \\
Vit4& 111.83 & 37.10 & 0.351 &0.37
& 0.547 & 1.8 & 18 &1.2 \\
Ce$_{68}$Al$_{10}$Cu$_{20}$Fe$_{2}$ & 32.20 & 11.80 & 0.337 &0.98
& 1.583 & 0.64 & 6.6 & 0\\
\end{tabular}
\end{ruledtabular}
\end{table*}

\subsection{\label{IIIB}Elastic modulus softening in LLRs}
\begin{figure}
 \centering
\includegraphics[width=8cm]{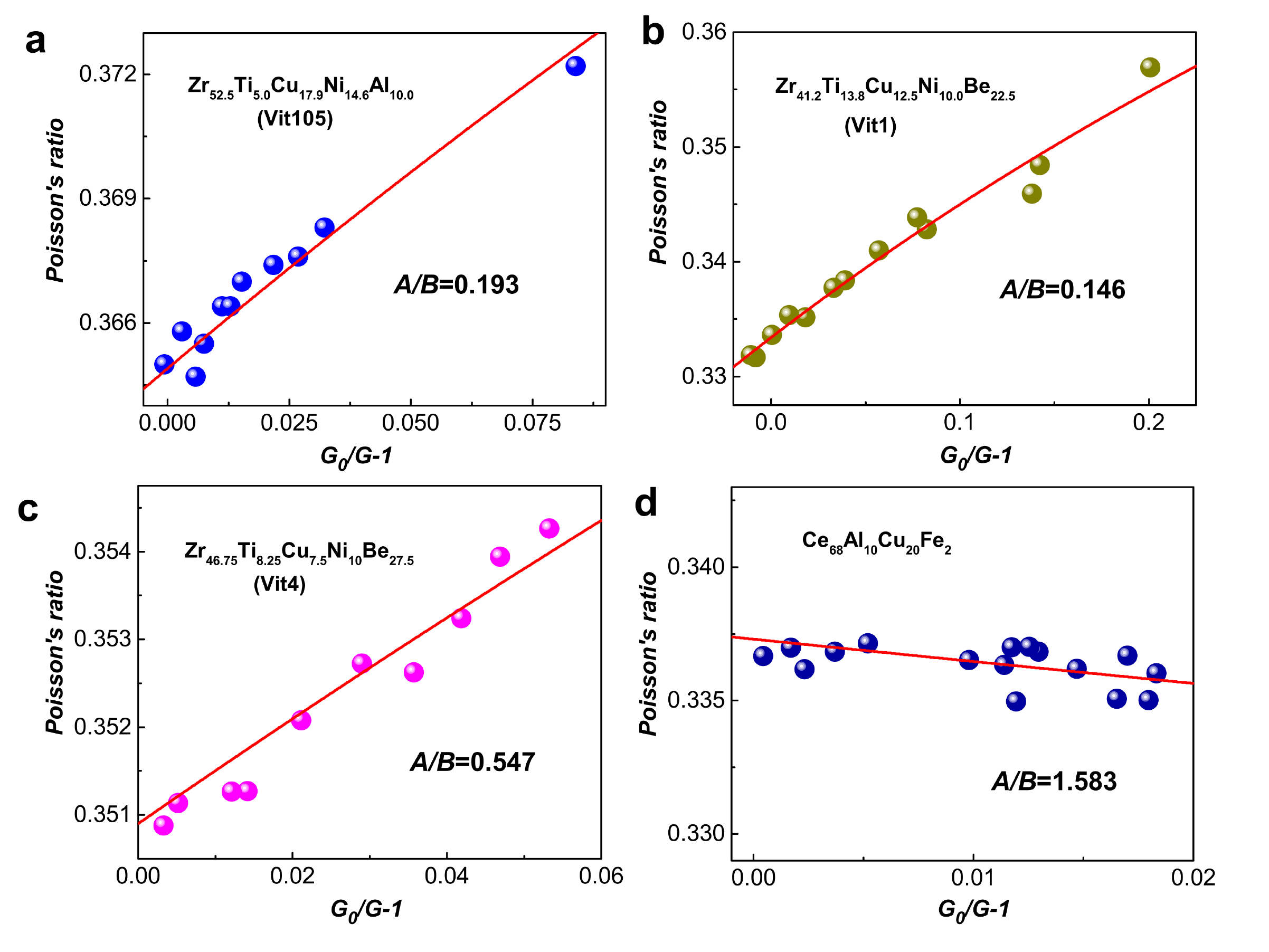}
\caption{\label{fig4}The plot of measured Poisson's ratio, $\nu$, versus $(G_{0}/G-1)$ upon annealing for different MGs: (a) Vit105; (b) Vit1; (c) Vit4; (d) Ce$_{68}$Al$_{10}$Cu$_{20}$Fe$_{2}$. All data were well fitted by Eq.\ref{eq6}, with the fitting values $A/B$ listed.  }
\end{figure}

 In general, it is difficult to directly measure the volume fraction of LLRs, $V_{f}$, in MGs. However, we noticed from Eq. \ref{eq2} that $V_{f}$ is related with the measured shear modulus upon annealing by: $V_{f}=(G_{0}/G-1)/B$. Taking this relation into Eq.\ref{eq5}, we obtain: 
                    \begin{equation}\label{eq6}
\nu=\frac{(1+\nu_{0})(G_{0}/G)-(1-2\nu_{0})[1+(A/B)(G_{0}/G-1)]}{2(1+\nu_{0})(G_{0}/G)-(1-2\nu_{0})[1+(A/B)(G_{0}/G-1)]}
                          \end {equation} 
  $G_{0}$, $\nu_{0}$ can be obtained by fitting the modulus data versus time, the ratio $A/B$ is the only adjustable parameter here. We plot the measured $\nu$ versus $(G_{0}/G-1)$ upon annealing for different MGs, as shown in Figure \ref{fig4}. Clearly, the experimental data can be well fitted by Eq.\ref{eq6}, verifying the validity of our theoretical model. Values of $A/B$ by fitting are shown in Figure \ref{fig4} and are found to have a compositional dependence. For Zr-based MGs (Vit105, Vit1 and Vit4), the ratio $A/B$ is less than 1, and for Ce-based MG, $A/B$ has a value larger than 1. As can be seen by Eq.\ref{eq3}, the ratio $A/B$ in fact reflects the relative size of $\xi$ and $\psi$, the softening degree of bulk modulus and shear modulus in LLRs, respectively. For $A/B<1$ as displayed by Zr-based MGs, we usually have $\xi<\psi$, or in other words, the shear-softening is dominated in LLRs, resulting in the increase of $\nu$ with $V_{f}$. While $A/B>1$ indicates a pressure-softening dominated process in LLRs, which leads to the decrease of $\nu$ with $V_{f}$ as displayed by the Ce-based MG. These results are in good agreement with our theoretical analysis. 

\begin{figure}
 \centering
\includegraphics[width=8cm]{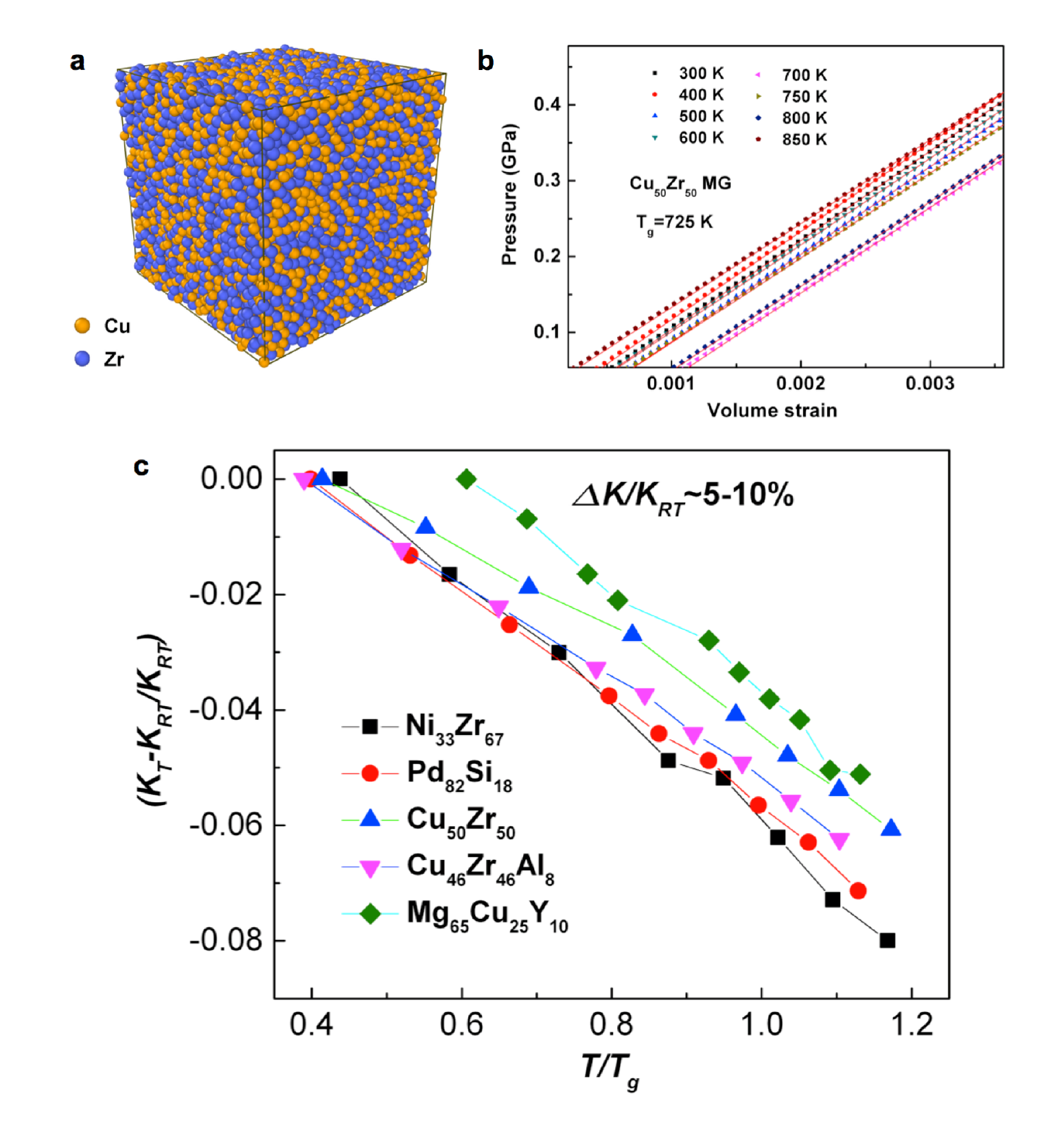}
\caption{\label{fig5} (a). The atomic configuration of Cu$_{50}$Zr$_{50}$ MG sample used in MD simulation. (b) The pressure-volume strain plot for Cu$_{50}$Zr$_{50}$ MG at different temperatures, from which the bulk modulus can be determined. (c) The relative change of bulk modulus $K$ from room temperature to well above $T_{g}$ for various MGs. }
\end{figure}

 It should be noted that the value of $A/B$ alone could not determine the exact softening degrees of $K$ and $G$ in LLRs. However, we could give a rough estimation on them if LLRs are considered to be in the supercooled liquid state, or as "residual liquidity" in MGs. According to previous studies, the relative change of bulk modulus is small ($<$10\% in general) compared to that of shear modulus in MGs, when the temperature increases from room temperature to the supercooled liquid regime. We also performed molecular simulations of five MGs with different potential functions and measured their bulk modulus change when the temperature increases from 300 K to temperatures well above their $T_{g}$, as shown in Figure \ref{fig5}. Strikingly, we found that the relative change of bulk modulus for all MGs are concentrated in a narrow range of 5-8\%. Given these results, we choose $\xi=$10\%. Together with the fitting value of $A/B$, we calculate the softening degree of shear modulus in LLRs,$\psi$, for different MGs, as listed in Table \ref{tableI}. One can see that $\psi$ largely varies with the MG composition and seems to positively correlate with the plasticity, ranging from 45\% for plastic Vit105 ($\epsilon_{p}\sim$5-6\%, $\epsilon_{p}$ is the final plastic strain) to 18\% for the less plastic Vit4 ($\epsilon_{p}\sim$1.5\% ) and 6.6\% for the completely brittle Ce-based MG (See Supplementary Figure 3). The correlation between  and the plasticity can be understood from the fact that LLRs with a relative large shear modulus softening are susceptible to inelastic deformation under a low shear stress, thus promote shear band formation at different sites, and result in a large plastic deformability ultimately. 

\section{Discussion}

\begin{figure}
 \centering
\includegraphics[width=8cm]{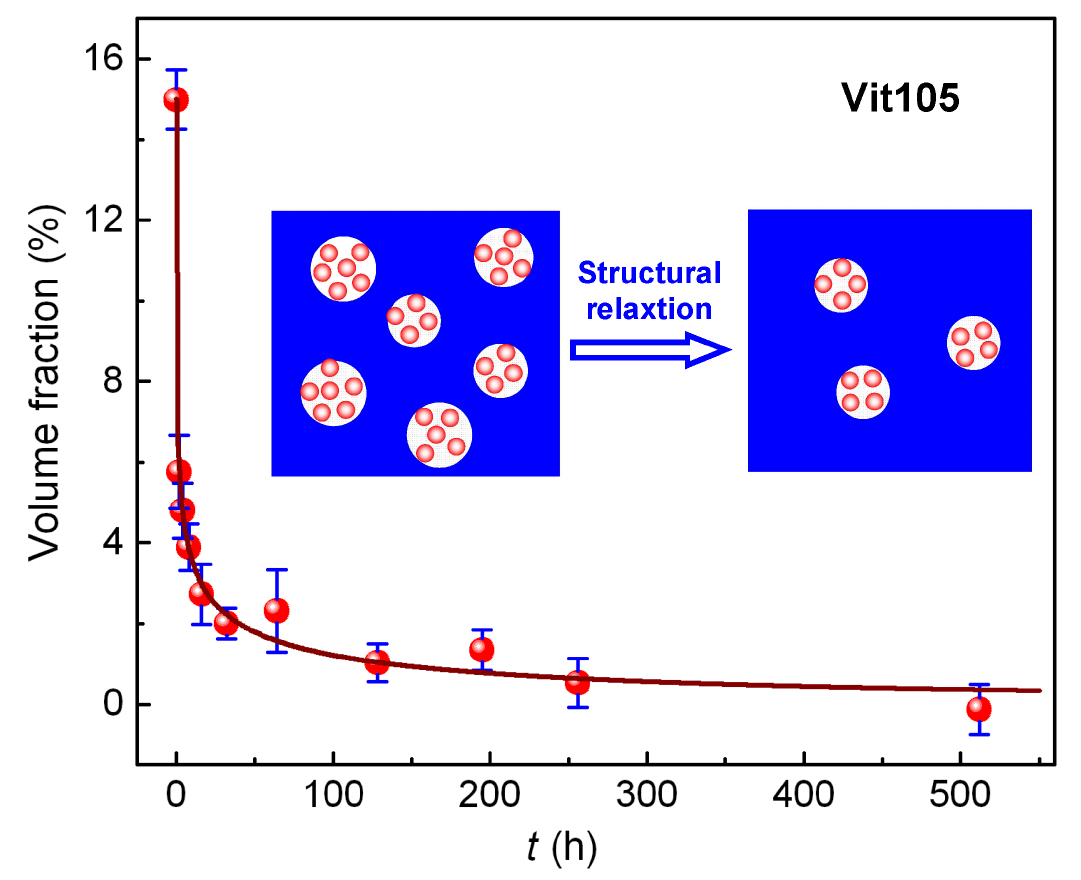}
\caption{\label{fig6} The calculated volume fraction of LLRs, $V_{f}$, versus with the annealing time $t$ for Vit105 based on the measured shear modulus $G$. The upper limit value of $\psi$ was used in the calculation.The inset schematic diagram illustrates the evolution of LLRs with time during structural relaxation. }
\end{figure}

Since LLRs are potential "flow units" responsible for macroscopic plastic flow and relaxation behavior in MGs[], it is important to investigate their evolutions dynamics with time. Based on the theoretical analysis and experimental results above, one can easily show that the evolution of LLR volume fraction with time upon annealing has the form: $V_{f}=V_{0}exp(-kt^{n})$, where $V_{0}$ is the volume fraction of LLRs in the as-cast state of MGs. This equation can well capture the experimentally determined $V_{f}$ at different annealing times (see Figure \ref{fig6}) from the measured shear modulus and the relation $V_{f}=(G_{0}/G-1)/B(\psi)$, where $\psi$ take the values listed in Table \ref{tableI}. The fitting values of $k$ and $n$ are the same as those obtained when fitting elastic moduli data. The form of $V_{f}(t)$ is similar to Johnson-Mehl-Avrami-Kohnogorov (JMAK) equation\cite{Fanfoni1998}, which usually describes the kinetic of isothermal solid-state transformation such as the precipitation of a crystalline phase from an amorphous matrix. The difference lies that the evolution of LLRs upon annealing is an annihilation process rather than a growth process, thus $V_{f}$ decreases monolithically with time and has the reverse form of JMAK equation. In addition, the fitting values of the exponent, $n$, is in the range $0-1$, which is also smaller than those reported in crystallization process($n\sim3-4$)\cite{MALEK199561}, which is associated with the three-dimensional nucleation and growth of crystalline phases in an amorphous matrix. To understand this, we developed a model by considering the reduction of LLRs both on their numbers and size during annealing (see Supplementary Text II). The theoretical analysis shows that the small values of $n$ ($\sim0-1$) mainly result from one-dimensional reduction of LLR size during annealing. Besides, the deviation of $n$ from $1$ is indicative of the spatial heterogeneity of LLR annihilation dynamics in MGs. 

   The established link between elastic properties and structural heterogeneity in MGs is helpful to understand many important correlations between elastic modulus and other properties reported in MGs. For example, the Poisson's ratio was shown to correlate well with the ductility or plasticity in MGs\cite{Lewandowski2005}, i.e. the larger plasticity favors the higher Poisson's ratio. Now, the correlation can be well understood from our analysis. From Eq. \ref{eq4}, one can see that the higher Poisson's ratio corresponds to the more significant structural heterogeneities of MGs, i.e. the larger volume fraction and the larger shear modulus softening of LLRs as well as the higher Poisson's ratio of the SLR matrix, factors all beneficial to the plasticity of MGs. It is also worth noting that the Poisson's ratio criterion for plasticity cannot be applied to MGs (such as the Ce-based MG in our present work) where the bulk modulus softening is much larger than shear modulus softening in LLRs, resulting in the decrease of Poisson's ratio with the volume fraction of LLRs. This may explain that the anomalous improved plasticity is achieved at lower Poisson's ratio reported in some Fe-based MGs\cite{poonapl}. 
   
   In addition, our present results could also be used to interpret the correlation between the Poisson's ratio in MGs and the fragility of their corresponding glass-forming liquids\cite{Novikov:2004aa}. It was once shown that the larger Possion's ratio or the ratio of shear modulus to bulk modulus in the glass often corresponds to the more fragile behavior of the liquid. Novikov et al\cite{Novikov:2004aa}interpreted the correlation from a non-ergodicity parameter, which determines the amplitude of fluctuations frozen at $T_{g}$ and thus relates the fast relaxation dynamics of a liquid to the ratio of shear modulus to bulk modulus in the glass state. Here, we could provide another interpretion on this correlation from the aspect of structural heterogeneities of glasses. Recent experimental studies\cite{ZhangboPRB} found that the fragility parameter, $m$, reflects the configuration change rate in the potential energy landscape of a liquid, and thus directly related with the temperature dependence of shear modulus softening around the glass transition. A collection of available experimental data on different kind of glasses\cite{ZhangboPRB,Wang2012487} have showed that the larger shear modulus softening rate at $T_{g}$ corresponds to the more fragile liquid or the larger value of $m$. Since LLRs can be regarded as the "residual liquid" in the structure of MG solids quenched from the liquid, it can be inferred that the liquid fragility is related to the shear modulus softening degree of LLRs as compared to SLRs in MGs. Meanwhile, provided that the volume fraction of LLRs is fixed, the large shear modulus softening degree of LLRs indicates a large value of Poisson's ratio from our current analysis. By this way, the fragility of a liquid can be linked to the Poisson's ratio of the glass solids quenched from the liquid. The details for origin of the correlation deserve a further study along this line.      
   
\section{\label{V}Summaries and Conclusions}

In summary, we established a quantitative link between overall elastic properties and local structural heterogeneities in MGs. General formulas for elastic moduli were derived from characteristics of structural hetergeneities from the Eshelby's theory and are verified by experimental results obtained from structural relaxation of various MGs. Particularly, we revealed that the decrease of Poisson's ratio upon annealing of MGs is associated with a much large shear softening over hydrostatic-pressure softening, and vice versa in local soft regions. Our results is helpful for understanding the Poisson's ratio criteria on the ductility and liquid fragility in MGs were also discussed. 

\begin{acknowledgments}
The work is supported by the Research Grant Council (RGC) of the Hong Kong government through the General Research Fund (GRF) with the account numbers CityU117612, 11209314, 9042066 and 9054013 and the NSF of China (51271195) and MOST 973 Program (No. 2015CB856800).
\end{acknowledgments}

\end{document}